\documentclass[aps,graphicx,showpacs,amsmath]{revtex4}
\usepackage{graphicx}
\usepackage{amssymb}

\def\ket{\rangle}
\begin{document}

\title{Realization of Quantum State Privacy Amplification in a Nuclear Magnetic
Resonance Quantum System\footnote{Liang Hao, ChuanWang and Gui Lu
Long, J. Phys. B: At. Mol. Opt. Phys. 43 (2010) 125502 (6pp),\\
doi:10.1088/0953-4075/43/12/125502}}
\author{Liang Hao$^{1}$, Chuan Wang$^{1,2}$ and Gui Lu Long$^{1,3}$\thanks{gllong@tsinghua.edu.cn}}
\address{$^{1}$Key Laboratory for Atomic and Molecular NanoSciences and Department of Physics,
Tsinghua University, Beijing 100084,
P. R. China\\
$^2$School of Science, Beijing University of Posts and
Telecommunications, Beijing 100876, China\\
$^3$Tsinghua National Laboratory for Information Science and
Technology, Beijing 100084, China}
\begin{abstract}
Quantum state privacy amplification (QSPA) is the quantum analogue
of classical privacy amplification. If the state information of a
series of single particle states has some leakage, QSPA reduces this
leakage by condensing the state information of two particles  into
the state of one particle. Recursive applications of the operations
will eliminate the quantum state information leakage to a required
minimum level.  In this paper, we report the experimental
implementation of  a quantum state privacy amplification protocol in
a nuclear magnetic resonance system. The density matrices of the
states are constructed in the experiment, and the experimental
results agree with theory well.
\end{abstract}

%\submitto{\JOB}
 \maketitle

%\date{\today}

\section{Introduction}
\label{s1}

The combination of quantum physics with communication gives rise to
quantum communication. The physics principles of quantum mechanics
offer the advantage of provable security, and higher capacity of
quantum communication over its classical counterparts. There are
various quantum communication tasks, such as quantum key
distribution where random keys are distributed among two users
separated at a distance \cite{Bennet84,Ekert91,Gisin02,Zhang05},
quantum secret sharing
\cite{Hillery99,Karlsson99,Bandyopadhyay00,Cleve99,Li04,Deng05}
where a secret is shared among several users and the users can read
out the shared secret only by cooperation, and quantum secure direct
communication
\cite{Long-Liu02,Beige,Felbinger02,Deng03,onetime,Yan04,Wang051}
where secret messages are transmitted directly over a quantum
channel. They serve the various needs of communication.

Under practical conditions, quantum communication is inevitably
affected by noise in quantum channels. It is hard  to distinguish
whether the errors are due to an eavesdropping behavior or the noise
in the channel itself. Therefore quantum communication over a noisy
channel  is not completely secure,  and some post-processing must be
done. For instance in quantum key distribution, privacy
amplification \cite{Bennett88} is performed to distill secure keys
from the less secure raw keys. Quantum privacy amplification
\cite{qpa1,qpa2,Deutsch96} is exploited for a sequence of entangled
Einstein-Podolsky-Rosen (EPR) pairs so as to obtain maximally
entangled pairs in protocols using EPR pairs.

In some quantum communication protocols, in  the end and
intermediate results of the quantum communication, there are a
sequence of single photons in nonorthogonal quantum states
\cite{onetime,zhangzj,zhangshou}, for instance  in states
 $\left|0\right\rangle$,
$\left|1\right\rangle$,
$\left|+x\right\rangle=(\left|0\right\rangle+\left|1\right\rangle)/\sqrt{2}$
and
$\left|-x\right\rangle=(\left|0\right\rangle-\left|1\right\rangle)/\sqrt{2}$,
where $|0\ket$ and $|1\ket$ represent the horizontal and vertical
polarization state of the photon respectively. These single photons
are sent from Bob to Alice. After eavesdropping check, Alice encodes
the bit values on the single photons by performing some unitary
operations, and then sends the encoded photons back to Bob. If the
information of the quantum states of the photons are leaked due to
either eavesdropping or channel noise, then one needs to reduce the
information leakage by some operation. The usual privacy
amplification used in QKD post-processing is not applicable, because
the single photons are in states in two conjugate basis, and they
cannot be added together in the manner used for the usual privacy
amplification. Quantum privacy amplification designed for the EPR
based protocols is not applicable to this case either, because the
single photons are not entangled states. For this purpose, quantum
state privacy amplification was proposed recently \cite{Deng06} and
it reduces the information leakage in the quantum states of single
photons. The essential idea of QSPA \cite{Deng06} is to perform
combined operations of controlled-NOT gates and Hadamard gates on
two qubits, and make a measurement on one qubit. Using the
measurement result, one can specify the state of the remaining
qubit. However the state of remaining particle is unknown to an
adversary, hence reduces his/her knowledge of the quantum state. In
this way, the information leakage is reduced.

 It is appealing to choose
the nuclear magnetic resonance (NMR) system to implement the QSPA
protocol. First, the NMR technique is a well-developed and
sophisticated technique. It has been a powerful tool for
experimental study of quantum information processing which  can
demonstrate the quantum manipulation of various quantum information
processing tasks. The essential features of quantum algorithms can
be demonstrated in such a quantum system, and the technique
developed in NMR system is also helpful in other candidate quantum
systems. For example, many quantum algorithms have been successfully
demonstrated in NMR quantum system
\cite{NMR1,NMR2,NMR3,NMR4,NMR5,NMR6}, and some quantum communication
protocols, such as teleportation \cite{nmrtele} and dense coding
\cite{nmrdense,nmrqss}, have been implemented in the NMR system.
Secondly, quantum gates are relatively easier to implement in NMR
system than in optical system. For transmission, it is no doubt that
optical system is the best candidate information carrier. However in
terms of gate operations, NMR system is easier because the quantum
gates can be realized by radio-frequency pulses and free evolutions.
Thus, demonstrating quantum algorithms in NMR enjoys the ease of
gate operations while testing the essential quantum operations. On
the other hand, there have been many efforts in building interfaces
between flying qubits and stationary qubits
\cite{inter1,inter2,inter3,inter4}. The combination of flying qubits
and stationary qubits may be a good candidate for quantum
information processing, especially for those involving both the
transmission and processing of qubits. If such interface could be
successfully build, nuclear spin qubit may well be a good candidate
of stationary qubit. In that case,  the flying photon qubit may
first be transferred to a stationary electron spin qubit, and then
further transferred to nuclear spin qubit.

This paper reports the experimental study of QSPA in a nuclear
magnetic resonance  quantum system. In this work, we have
experimentally implemented the QSPA protocol in a nuclear magnetic
resonance system, and all the quantum operations needed in the QSPA
are demonstrated. Density matrices of states during the QSPA were
constructed. The experimental results agree with theory well.

\section{The Principle of Quantum State Privacy Amplification}
\begin{figure}[!h]
\centering
\includegraphics[height=7.5cm,angle=90]{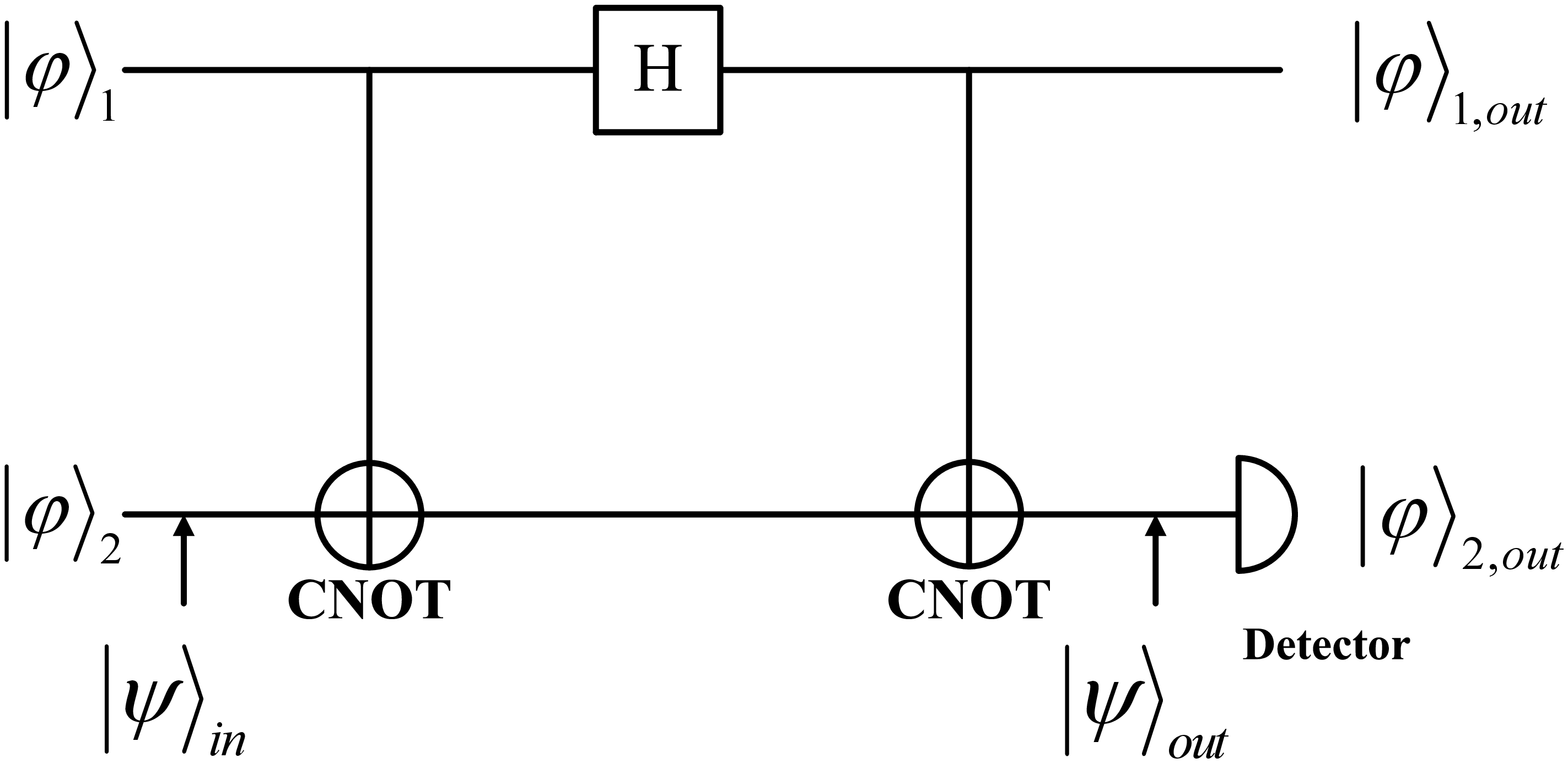}
\caption{Circuit of the CHC-QSPA protocol. $\left| \varphi_1
\right\rangle$ and $\left| \varphi_2 \right\rangle$ denotes the
states of the two qubits, respectively. The target qubit is measured
at the end so as to incorporate the state information of  the second
qubit into the first qubit.} \label{f1}
\end{figure}

Firstly, we briefly describe the QSPA protocol, for details  see
Ref. \cite{Deng06}. It is the quantum analogue of classical privacy
amplification. In classical cryptography, if a sequence key of $n$
bits has some leakage to the outside, the legitimate users can use
the privacy amplification to reduce this information leakage. The
common privacy amplification protocol \cite{Bennett88} uses the
parities of $m$ partitions of subsets of the original raw key with
some permutations. Thus, instead of using the original $n$-bits key,
the legitimate users use
 $m$ bits of parities as the new key. Usually, $m$ is less than
$n$, the privacy of the key has been amplified, and hence has better
security.  The task of QSPA is as follows. Two single-qubit states,
\begin{eqnarray}
 \left| \varphi \right\rangle
_1 & = & a_1 \left| 0 \right\rangle  + b_1 \left| 1 \right\rangle,\\
\left| \varphi  \right\rangle _2 & = &a_2 \left| 0 \right\rangle  +
b_2 \left| 1 \right\rangle,
\end{eqnarray}
where the coefficients $a_1$, $b_1$, $a_2$ and $b_2$ satisfy the
normalization requirement,
\begin{eqnarray}
\left| {a_1 } \right|^2  + \left| {b_1 } \right|^2  = \left| {a_2 }
\right|^2  + \left| {b_2 } \right|^2  = 1,
\end{eqnarray}
are known to the legitimate users, and however have an information
leakage to an adversary Eve. The task of QSPA is to reduce this
state information leakage. Because the single-qubit states are
usually not orthogonal, the approach to take the parity is not
applicable. Instead, the quantum state privacy amplification
protocol in Ref.\cite{Deng06} uses  two controlled-not (CNOT) gates
and a Hadamard (H) gate, which may be simply called the CHC
operation (for simplicity, we call this QSPA protocol as CHC-QSPA
protocol hereafter). The schematic circuit is shown in Fig.\ref{f1}.
The initial state $\left|\psi\right\rangle_{in}$ of the QSPA is the
product of two single photon states,
\begin{eqnarray}
\left|\psi\right\rangle_{in} = \left| \varphi  \right\rangle _1
\otimes \left| \varphi  \right\rangle _2.
\end{eqnarray}

After the CHC operation, the  state of the joint system is changed
to
\begin{eqnarray}
\left| \psi  \right\rangle _{out}  = {1 \over {\sqrt 2 }}\{ (a_1 a_2
+ b_1 b_2 )\left| 0 \right\rangle _1  + (a_1 b_2  - b_1 a_2 )\left|
1 \right\rangle _1 \} \left| 0 \right\rangle _2  \nonumber \\
+ {1 \over {\sqrt 2 }}\{ (a_1 a_2  - b_1 b_2 )\left| 1 \right\rangle
_1  + (a_1 b_2  + b_1 a_2 )\left| 0 \right\rangle _1 \} \left| 1
\right\rangle _2.\nonumber \\\label{e1}
\end{eqnarray}
Then one measures the second qubit in the $\sigma_z$ basis. If
$\left| \varphi \right\rangle _{2,out} =\left| 0 \right\rangle$ is
obtained, the state of control qubit is
\begin{eqnarray}
\left| \varphi \right\rangle _{1,out} = (a_1 a_2 + b_1 b_2 )\left| 0
\right\rangle _1  + (a_1 b_2  - b_1 a_2 )\left| 1 \right\rangle _1.
\end{eqnarray}
Otherwise, the first qubit state is
\begin{eqnarray}
\left| \varphi \right\rangle _{1,out} =  (a_1 a_2  - b_1 b_2 )\left|
1 \right\rangle _1  + (a_1 b_2  + b_1 a_2 )\left| 0 \right\rangle
_1.
\end{eqnarray}
In this way, no matter what the measurement result of second qubit
is, the state information of two qubits is concentrated on the first
one. So the privacy of the state is amplified.

\begin{table}[!h]
\begin{center}
\begin{tabular}{c|cccc}\hline
& \multicolumn{4}{c}{$\varphi_{1}$}\\ \cline{2-5} $\varphi_{2 }$&
$\vert +z\rangle $ & $\vert -z\rangle $ & $
\vert +x\rangle $ & $\vert -x\rangle$\\
\hline
 $\vert +z\rangle $ & $\vert 0\rangle$ & $\vert 1\rangle $ & $\vert -x\rangle$  & $\vert +x\rangle$ \\
$\vert -z\rangle $ & $\vert 1\rangle $ & $\vert 0\rangle $ & $\vert
+x\rangle$ &
$\vert -x\rangle$ \\
$\vert +x\rangle $ & $\vert +x\rangle$ & $\vert -x\rangle$ &
$\vert 0\rangle $ & $\vert 1\rangle $ \\
$\vert -x\rangle $ & $\vert -x\rangle$ & $\vert +x\rangle$ &
$\vert 1\rangle $ & $\vert 0\rangle $\\
\hline
\end{tabular}\label{Table1}
\caption{The truth table of the output state  for the control qubit
$\left| \varphi \right\rangle _{1,out}$, when the measurement result
of the target qubit is $\left| 0 \right\rangle $.  $\left| \varphi
\right\rangle _1$ and $\left| \varphi  \right\rangle _2$ are the
input states of the control and target
qubit,respectively.}\label{t1}
\end{center}
\end{table}

\begin{table}[!h]
\begin{center}
\begin{tabular}{c|cccc}\hline
 & \multicolumn{4}{c}{$\varphi_1$}\\
 \cline{2-5}
$\varphi_2$ & $\vert +z\rangle $ & $\vert -z\rangle $ & $ \vert
+x\rangle $ & $\vert -x\rangle$ \\
\hline $\vert +z\rangle $ & $\vert 1\rangle $ & $\vert 0\rangle $ &
$\vert +x\rangle$ &
$\vert -x\rangle$ \\
$\vert -z\rangle $ & $\vert 0\rangle $ & $\vert 1\rangle $ & $\vert
-x\rangle$ &
$\vert +x\rangle$ \\
$\vert +x\rangle $ & $\vert +x\rangle$ & $\vert -x\rangle$ &
$\vert 0\rangle $ & $\vert 1\rangle $ \\
$\vert -x\rangle $ & $\vert -x\rangle$ & $\vert +x\rangle$ &
$\vert 1\rangle $ & $\vert 0\rangle $\\
\hline
\end{tabular}\label{t2}
\caption{The truth table of the output state of the control qubit
$\left| \varphi \right\rangle _{1,out}$, when the measurement result
of the target qubit is $\left| 1 \right\rangle $.  $\left| \varphi
\right\rangle _1$ and $\left| \varphi  \right\rangle _2$ are the
input states of the control and target qubit,respectively.}
\end{center}
\end{table}

We take the quantum one-time-pad protocol \cite{onetime} as an
example, where the single photon states are $|\pm x\ket$ and $|\pm
z\ket$ respectively.  The truth table of the output state of first
qubit $\left| \varphi \right\rangle _{1,out}$ is shown in  Tables
\ref{t1} and \ref{t2} for the measurement results of the target
qubit $\left| 0 \right\rangle $ and $\left| 1 \right\rangle$,
respectively. When $\left| \varphi \right\rangle _1
=\left|0\right\rangle$, $\left| \varphi \right\rangle _2
=\left|1\right\rangle$, after the CHC operation, the output state is
\begin{eqnarray}
\left| \psi \right\rangle _{out}
=\frac{1}{\sqrt{2}}(\left|1\right\rangle_1
\left|0\right\rangle_2+\left|0\right\rangle_1
\left|1\right\rangle_2).
\end{eqnarray}
If the measurement result of the target qubit $\left| \varphi
\right\rangle _{2,out}=\left|0\right\rangle$, the final state of the
control qubit $\left| \varphi \right\rangle _{1,out}=
\left|1\right\rangle$. So the privacy of the input state
$\left|0\right\rangle_1 \left|1\right\rangle_2$ is concentrated on
the final state of the control qubit $ \left|1\right\rangle_1$. If
$\left| \varphi \right\rangle _{2,out}=\left|1\right\rangle$ is
obtained, the initial key $\left|0\right\rangle_1
\left|1\right\rangle_2$ is compressed to the condensed key $
\left|0\right\rangle_1$. It is obvious that the condensed key
depends not only on the input states of two single photons, but also
on the result of the measurement on the target qubit.

Suppose an adversary Eve knows the complete information of the
control qubit but nothing about the target qubit. She can only guess
the input state of the target qubit to deduce the condensed key. In
most quantum communication protocols, the single photons prepared by
the legitimate user are in one of the four states
$\left|0\right\rangle$, $\left|1\right\rangle$,
$\left|+x\right\rangle$, $\left|-x\right\rangle$ randomly.  Even if
Eve knows completely the information of both qubits, which is a
small probability event\cite{Deng06}, she can not deduce the
condensed state with certainty,  because the final state of the
control qubit $\left| \varphi \right\rangle _{1,out}$ has two
possible results even though the input state $\left| \psi
\right\rangle _{in}$ is fixed. The output state depends on the
measurement result of the target qubit, according to the truth
tables, when the input state $\left| \varphi
\right\rangle_2=\left|0\right\rangle$.

In practice, the process of QSPA can be used repeatedly to get more
secure quantum state by using the retained qubit from the last round
as a control and choosing a third qubit from the sequence as the
target. The more one repeats the QSPA operations, the lower the
state information leakage.

\section{NMR realization}
The QSPA protocol was experimentally realized in NMR in a sample of
Carbon-13 labeled chloroform ($^{13}$CHCL$_{3}$) which was dissolved
in d6 acetone. The experiments were done at $22^\circ C$ with a
Bruker Avance III 400 MHz spectrometer. We assign the $^{13}$C as
the control qubit, and the $^{1}$H as the target qubit, here the two
qubits are denoted as C1 and H2. With the convention that magnetic
field is along the $z$-axis, we define the spin up $\left|\uparrow
\right\rangle$ as $\left|0\right\rangle$ state and spin down
$\left|\downarrow \right\rangle$ as $\left|1\right\rangle$ state,
which correspond to the low-frequency part and high-frequency part
of the spectrum respectively. The system Hamiltonian is
\begin{eqnarray}
H_{NMR}=-\pi\nu_{1}\sigma_{z}^{1}-\pi\nu_{2}\sigma_{z}^{2} +
\frac{1}{2}\pi J_{12}\sigma_{z}^{1}\sigma_{z}^{2},
\end{eqnarray}
where $\nu_{1}$ and $\nu_{2}$ are the resonance frequencies for C1
and H2, respectively. The scalar coupling between two spins $J_{12}$
has been measured to be $215$Hz.

 The pseudopure state
$\left|0\right\rangle_1\left|0\right\rangle_2 $
 is prepared from the thermal equilibrium state of the two qubits system
 using the spatial averaging method \cite{NMR2,NMR5}, where  the pulse
 sequence is
$[\theta]_{x}^{2}\rightarrow[grad]_{z}\rightarrow[-\pi/4]_{x}^{2}\rightarrow[1/2J]\rightarrow[\pi/4]_{y}^{2}\rightarrow[grad]_{z}$,
where $[grad]_z$ is the gradient field.  The state evolution under
these pulses is
\begin{eqnarray}
& & \gamma_{C}I_z ^1  + \gamma_{H}I_z ^2 \nonumber\\
      &\left[\theta\right]_x ^2\Rightarrow
 & \gamma_{C}(I_z ^1+2I_z ^2)  - \sqrt{1-\frac{4\gamma_{C}^{2}}{\gamma_{H}^{2}}}I_y ^2 \nonumber\\
 &\left[ grad \right]\Rightarrow
& \gamma_{C}(I_z ^1  + 2I_z ^2)  \nonumber\\
 & \left[
\frac{\pi}{4}\right]_{-x} ^2\Rightarrow & \gamma_{C}(I_z ^1 + {\sqrt 2} I_z ^2  + {\sqrt 2}I_y ^2) \nonumber\\
 &\left[
\frac{1}{2J_{12}}\right]\Rightarrow& \gamma_{C}(I_z^1 + {\sqrt 2} I_z ^2 - {2\sqrt 2 }I_z ^1 I_x ^2) \nonumber\\
 &\left[
\frac{\pi}{4}\right]_{y} ^2 \Rightarrow & \gamma_{C}(I_z ^1  +
I_z^2+I_x^2-2I_z^1I_x^2+2I_z^1I_z^2)\nonumber\\
&\left[ grad \right] \Rightarrow{\;\;\;\;\;\;\;\;\;}&
2\gamma_{C}[(\frac{1}{2}+I_z ^1)(\frac{1}{2}+ I_z ^2)-\frac{1}{4}],
\end{eqnarray}
where $\left[\theta\right]_{\alpha}^{k}$ is defined as the rotation
of spin $k$ through angle $\theta$ about $\hat{\alpha}$-axis.
$\left[\tau\right]$ is the free evolution of the system for $\tau$
time interval, and $[grad]_{z}$ denotes a gradient pulse along the
$\hat{z}$-axis. Accordingly,
\begin{eqnarray}
\cos\theta=2\gamma_{C}/\gamma_{H},
\end{eqnarray}
where $\gamma_{C}$ and $\gamma_{H}$ are the gyromagnetic ratios of
spin $^{13}$C and proton $^1$H, respectively.

\begin{figure}[!h]
%\centering
\includegraphics[height=7.5cm,angle=-90]{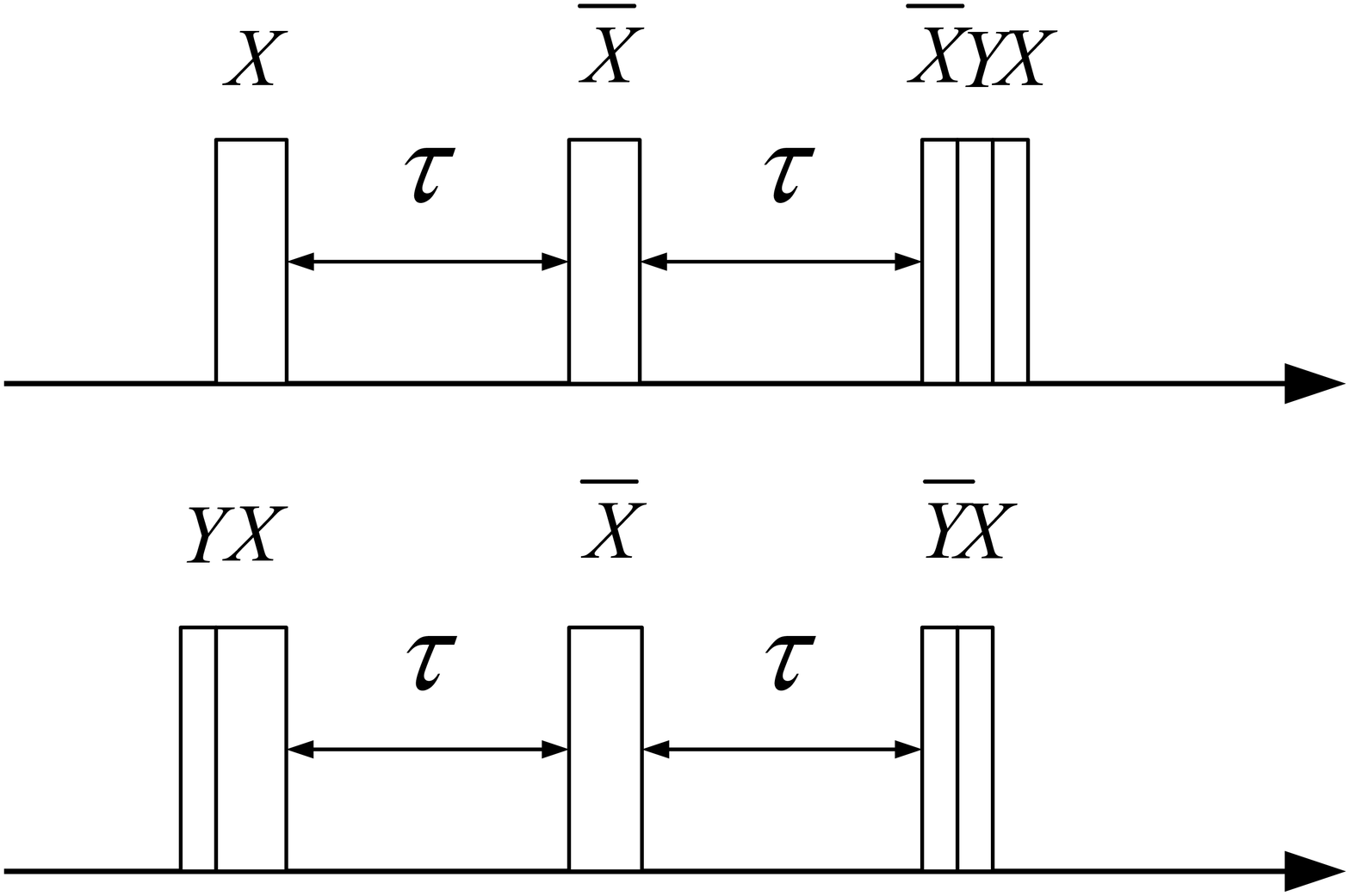}
\caption{Optimal NMR pulse sequence for the controlled-NOT gate. The
wide and narrow boxes denote $\pi$ and $\pi/2$ pulses, respectively.
the upper and lower lines are for the control and target qubits
respectively. X and Y denote the axes along which the pulses are
applied, and the overbars indicate the opposite direction. The time
period $\tau$, during which no pulses are applied, is set equal to
$1/4J_{12}$. The order of pulses are  from the left to the right.}
\label{f2}
\end{figure}

\begin{figure}[!h]
\begin{center}
\includegraphics[width=9cm]{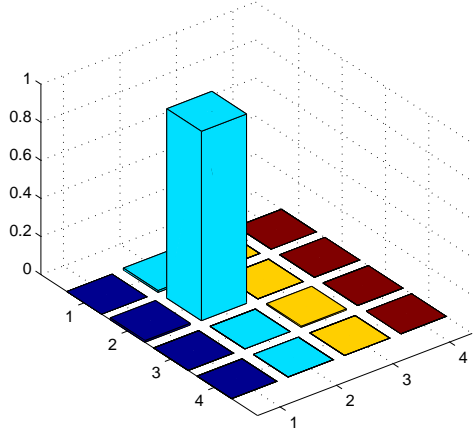}
\caption{The density matrix of initial state
$\left|\psi\right\rangle_{in}=\left | 0 \right \rangle_1 \left | 1
\right \rangle_2$.} \label{f3}
\end{center}
\end{figure}

\begin{figure}[!h]
\begin{center}
\includegraphics[width=9cm]{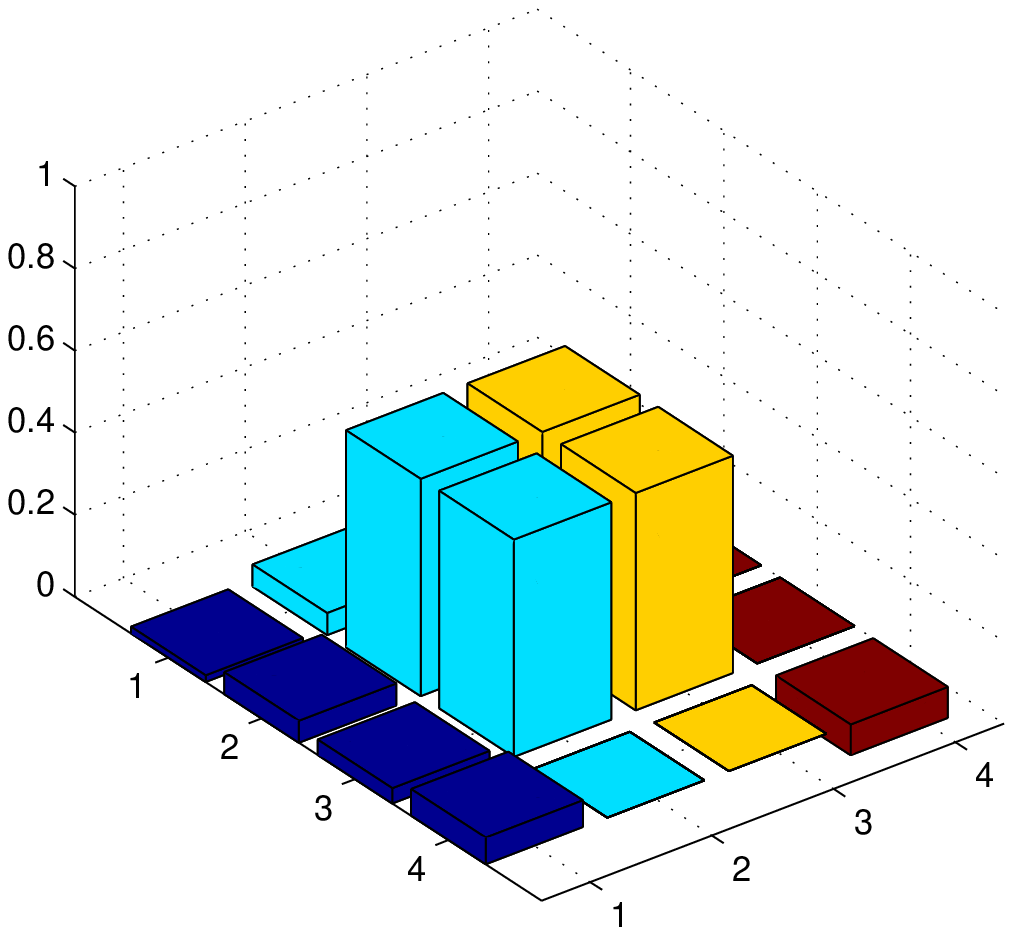}
\caption{The density matrix of outcome state
$\left|\psi\right\rangle_{out}$ after the QSPA operation on state
$\left | 0 \right \rangle_1 \left | 1 \right \rangle_2$.} \label{f4}
\end{center}
\end{figure}

The pulse sequence of the QSPA operation in the CHC-QSPA protocol is
found to be
%\begin{widetext}
%\begin{center}
\begin{eqnarray}
&&[\frac{\pi}{2}]_{y}^{2}\rightarrow[\frac{1}{2J_{12}}]\rightarrow[\pi]_{-y}^{2}\rightarrow[\frac{\pi}{2}]_{x}^{2}\nonumber\\
&&\rightarrow[\frac{\pi}{2}]_{z}^{1,2}
\rightarrow[\frac{\pi}{2}]_{y}^{1}
\rightarrow[\pi]_{-x}^{1}\rightarrow[\frac{\pi}{2}]_{y}^{2}\nonumber\\
&&\rightarrow[\frac{1}{2J_{12}}]
\rightarrow[\pi]_{-y}^{2}\rightarrow[\frac{\pi}{2}]_{x}^{2}\rightarrow[\frac{\pi}{2}]_{z}^{1,2}.\nonumber\\
\end{eqnarray}
The optimized CNOT gate  pulse sequence is shown in Fig.\ref{f2}.
The free evolution $[\frac{1}{2J_{12}}]$ is realized by using a free
time delay $\tau=1/4J_{12}$ separated by a pair of $\pi$ pulses in
the opposite directions, and these pulses can  average out the
effects of free Hamiltonian evolution of two spins $\sigma_z^1$ and
$\sigma_z^2$. In this way, we can also reduce the error
accumulations caused by imperfect calibration of the $\pi$ pulses.
The H gate is implemented by a $\pi/2$ pulse along $\hat{y}$-axis,
followed by a $\pi$ pulse along $\hat{-x}$-axis.

%\begin{widetext}
\begin{center}
\begin{figure}[!h]
\includegraphics[width=5.5in]{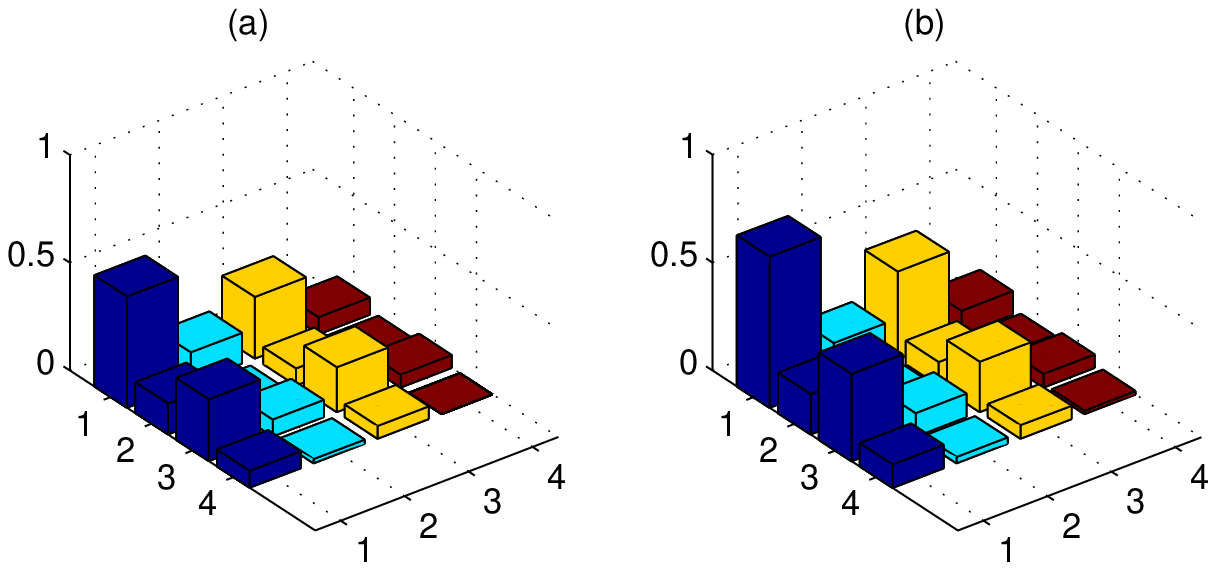}
\caption{The density matrix of initial state
$\left|\psi\right\rangle_{in}=\left( {{{\sqrt 3 } \over 2}\left| 0
\right\rangle_1  + {1 \over 2}\left| 1 \right\rangle_1 }
\right)\left( {\cos 15^o \left| 0 \right\rangle_2 + \sin 15^o \left|
1 \right\rangle_2 } \right)$: (a) experiment result; (b) theoretical
prediction.} \label{f5}
\end{figure}
\end{center}

\begin{center}
\begin{figure}[!h]
\includegraphics[width=5.5in]{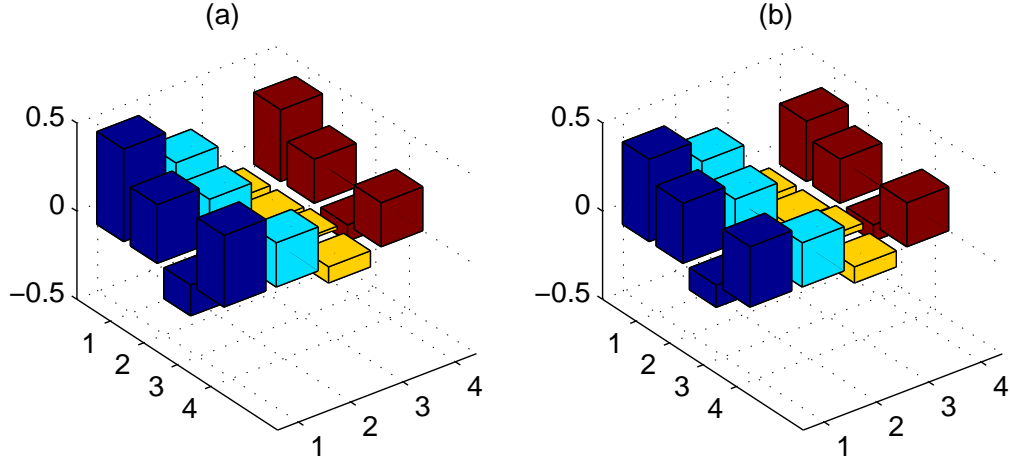}
\caption{The density matrix of outcome state
 after the CHC-operation
$\left( {{{\sqrt 3 } \over 2}\left| 0 \right\rangle_1  + {1 \over
2}\left| 1 \right\rangle_1 } \right)\left( {\cos 15^o \left| 0
\right\rangle_2 + \sin 15^o \left| 1 \right\rangle_2 } \right)$: (a)
experimental result; and (b) theoretical result.} \label{f6}
\end{figure}
\end{center}

%$[\frac{\pi}{2}]_{y}^{2}\rightarrow[\frac{1}{2J_{12}}]\rightarrow[\pi]_{-y}^{2}\rightarrow[\frac{\pi}{2}]_{x}^{2}\rightarrow[\frac{\pi}{2}]_{z}^{1,2}\rightarrow[\frac{\pi}{2}]_{y}^{1}\rightarrow[\pi]_{-x}^{1}\rightarrow[\frac{\pi}{2}]_{y}^{2}\rightarrow[\frac{1}{2J_{12}}]
%\rightarrow[\pi]_{-y}^{2}\rightarrow[\frac{\pi}{2}]_{x}^{2}\rightarrow[\frac{\pi}{2}]_{z}^{1,2}$.
We have implemented the QSPA protocol for two kinds of initial
states,  $\left|\psi\right\rangle_{in}=\left | 0 \right \rangle_1
\left | 1 \right \rangle_2$, and
$\left|\psi\right\rangle_{in}=\left( {{{\sqrt 3 } \over 2}\left| 0
\right\rangle_1  + {1 \over 2}\left| 1 \right\rangle_1 }
\right)\left( {\cos 15^o \left| 0 \right\rangle_2 + \sin 15^o \left|
1 \right\rangle_2 } \right)$. The state
$\left|0\right\rangle_1\left|1\right\rangle_2$ can be obtained by
rotating the second spin $\pi$ about $\hat{y}$-axis from the
pseudopure state $\left|0\right\rangle_1\left|0\right\rangle_2$. The
experimental density matrix for initial state  $\left | 0 \right
\rangle_1 \left | 1 \right \rangle_2$ is shown in Fig.\ref{f3}.
After the CHC-operation, the outcome state is
$\left|\psi\right\rangle_{out}={1 \over {\sqrt 2 }}(\left| {01}
\right\rangle  + \left| {10} \right\rangle )$, and the experimental
density matrix was reconstructed by state tomography technique
\cite{NMR7,NMR8} and is shown in Fig. \ref{f4}. The density matrices
agree with theoretical predictions very well.

For quantum communication, the case with  state
$\left|\psi\right\rangle_{in}=\left( {{{\sqrt 3 } \over 2}\left| 0
\right\rangle_1  + {1 \over 2}\left| 1 \right\rangle_1 } \right)$
$\left( {\cos 15^o \left| 0 \right\rangle_2 + \sin 15^o \left| 1
\right\rangle_2 } \right)$ is more interesting because it represents
the more general nonorthogonal state case.   The state $\left(
{{{\sqrt 3 } \over 2}\left| 0 \right\rangle_1 + {1 \over 2}\left| 1
\right\rangle_1 } \right)\left( {\cos 15^o \left| 0 \right\rangle_2
+ \sin 15^o \left| 1 \right\rangle_2 } \right)$ can be prepared by
applying
$[\frac{2\pi}{3}]_{y}^{1}\rightarrow[\frac{\pi}{3}]_{y}^{2}$ to the
pseudopure state. The density matrices of the initial and output
state are shown in Fig.\ref{f5} and \ref{f6} respectively. For
comparisons, we have also shown the theoretical density matrices for
these states.  The output state has a more complicated form for this
initial state, and it is $\left|\psi\right\rangle_{out}=\left(
0.6830 \left| {00} \right\rangle+0.5000\left| {01} \right\rangle -
0.1830\left| {10} \right\rangle+0.5000\left| {11} \right\rangle
\right)$. Fig.\ref{f6}(a) shows the density matrix of the outcome
state.  The agreement between theory and experiment is good.

\section{Summary}

In conclusion, we have constructed the pulse sequence for the QSPA
operation in the CHC-QSPA protocol. The results of the experiments
agree with the theoretical predictions well both in  the case with
computational basis states and the general case of nonorthogonal
states. The output carries the state information of two qubits in
the QSPA. With the information about the measurement result and the
information of the two input qubits, a legitimate user knows the
state of qubit after QSPA, whereas an illegal user lacks the proper
information of all the qubits involved, and hence loses his/her
knowledge about the quantum state after the QSPA. In this way, the
quantum state information leakage is reduced. The present experiment
clearly demonstrated the protocol.

This work is supported by the National Natural Science Foundation of
China Grant No.  10874098,  the National Basic Research Program of
China (2006CB921106,  2009CB929402).

\end{document}